# All-Metallic Electrically-Gated 2H-TaSe$_2$ Switches and Logic Circuits

J. Renteria[1], R. Samnakay[2], C. Jiang[1], T.R. Pope[3], P. Goli[2], Z. Yan[1], D. Wickramaratne[4], T.T. Salguero[3], A.G. Khitun[1,2], R.K. Lake[4] and A.A. Balandin[1,3,×]

[1]Nano-Device Laboratory, Department of Electrical Engineering, Bourns College of Engineering, University of California – Riverside, Riverside, California 92521 USA

[2]Materials Science and Engineering Program, Bourns College of Engineering, University of California – Riverside, Riverside, California 92521 USA

[3]Department of Chemistry, University of Georgia, Athens, Georgia 30602 USA

[4]Laboratory for Terascale and Terahertz Electronics, Department of Electrical Engineering, Bourns College of Engineering, University of California – Riverside, Riverside, California 92521 USA

## Abstract

We report the fabrication and performance of *all-metallic* three-terminal devices with tantalum diselenide thin-film conducting channels. For this proof-of-concept demonstration, the layers of 2H-TaSe$_2$ were exfoliated mechanically from single crystals grown by the chemical vapor transport method. Devices with nanometer-scale thicknesses exhibit strongly non-linear current-voltage characteristics, unusual optical response, and electrical gating at room temperature. We have found that the drain-source current in thin-film 2H-TaSe$_2$–Ti/Au devices reproducibly shows an abrupt transition from a highly resistive to a conductive state, with the threshold tunable via the gate voltage. Such current-voltage characteristics can be used in principle for implementing *radiation-hard* all-metallic logic circuits. These results may open new application space for thin films of *van der Waals* materials.

× Corresponding author: balandin@ee.ucr.edu; http://ndl.ee.ucr.edu/

## I.      Introduction

Implementation of transistors, diodes and other circuit elements with metallic materials rather than semiconductors would offer a number of potential benefits including inherent radiation hardness and low cost. There have been a number of proposals and attempts to implement all-metallic switches and circuits ranging from metal dot single electron transistors[1-3] to metallic carbon nanotube devices,[4-6] metallic nanowire transistors[7] and all-metallic spin transistors.[8,9] To date the most successful demonstration of an electrically-gated metallic channel device is a graphene field-effect transistor (FET) type device.[10-12] Graphene devices have high mobility[11,12] high saturation velocity,[13] low flicker noise[14,15] and can be gated at room temperature (RT).[10-12] The V-shaped transconductance characteristics of graphene devices and their low $1/f$ noise level offer increased functionalities for signal processing ($f$ is the frequency).[15-17] However, the absence of a band gap, the high carrier concentration, and the high mobility in graphene result in substantial leakage currents. In addition, graphene-based devices cannot be switched off, and the single-atom thickness of graphene leaves it susceptible to radiation damage.[18]

In this paper, we demonstrate that electrically-gated devices with strongly non-linear current-voltage (I-V) characteristics can be based on thin films of metallic *van der Waals* materials. The drain-source current in such devices reveals an abrupt transition from a highly resistive to a conductive state with the threshold tunable via the gate voltage. We further show that such I-V characteristics can be used to implement *all-metallic* logic circuits that are potentially *radiation-hard*. These devices are based on tantalum diselenide ($TaSe_2$) thin-film conducting channels.

## II.      Material Synthesis and Device Fabrication

$TaSe_2$ is representative of the layered transition metal dichalcogenides, $MX_2$, where M=Mo, W, Nb, Ta or Ti and X=S, Se or Te.[19] Some of these materials manifest charge density wave (CDW) phenomena in the temperature range from ~100 K to RT. We have reported previously that FETs with $TiTe_2$ channels can show unusual non-linear I-V characteristics.[20] Unlike $MoS_2$, which is a semiconductor with a bulk band gap energy of $E_G$=1.29 eV (1.9 eV for a monolayer),[21] 2H-$TaSe_2$ is a metal (or semi-metal) with no band gap.[22] The crystal structures of $TaSe_2$ contain Ta in trigonal prismatic coordination within Se-Ta-Se layers. The weak interlayer van der Waals bonding leads to various TaSe$_2$ polytypes, which differ simply in the relative orientations of the layers and their stacking arrangements. In this work, we focus on 2H-TaSe$_2$. As illustrated in Figure 1, the structure of the 2H polytype has a two-layer repeat pattern (AcA BcB); the view down the c-axis clearly shows how the trigonal prismatic units are rotated 60$^o$ with respect to each other.[23]

High purity and crystallinity are essential to ensure desirable electron transport characteristics. Thus, the 2H-TaSe$_2$ thin films used in this work were exfoliated mechanically from single crystals grown by the chemical vapor transport (CVT) method. Preparative details and analysis data can be found in a recent report.[24] Representative images of 2H-TaSe$_2$ crystals are shown in Figure 1: an as-grown, metallic gray cluster of crystals in a vial, and a scanning electron microscopy (SEM) image highlighting an individual 2H-TaSe$_2$ crystal. Lateral dimensions ranged from 200 to 800 μm and thicknesses were typically 5 to 30 μm.

[Figure 1]

These crystals of 2H-TaSe$_2$ were exfoliated and transferred onto Si/SiO$_2$ substrates following the standard "graphene-like" approach.[10-12] The thickness $H$ of the films ranged from a few tri-layers to $H \approx 40$ nm. Micro-Raman spectroscopy (Renishaw InVia) was used to verify the crystallinity, polytype and thickness of the flakes after exfoliation. It was performed in a backscattering configuration and $\lambda$=633-nm excitation laser. The excitation laser power was limited to less than 0.5 mW to avoid local heating. Figure 2 shows informative bands at 207 – 210 cm$^{-1}$ (E$^1_{2g}$) and 233 – 235 cm$^{-1}$ (A$_{1g}$), consistent with previous reports for bulk 2H-TaSe$_2$.[24, 25] The band at 300 cm$^{-1}$ corresponds to the Si/SiO$_2$ substrate.

[Figure 2]

The devices with 2H-TaSe$_2$-film channels were fabricated using electron beam lithography (LEO SUPRA 55) for source and drain electrode patterning and electron-beam evaporation (Temescal BJD-1800) for metal deposition. Si substrates with 300-nm thick SiO$_2$ layers were spin coated (Headway SCE) and baked consecutively with two positive resists: MMA (methyl methacrylate) and then PMMA (polymethyl methacrylate). After patterning of the source and drain contacts to the 2H-TaSe$_2$ flakes, the metal layers of Ti/Au (10-nm / 70-nm) were deposited to complete the devices. The heavily doped Si/SiO$_2$ wafer served as a back gate. Most of the devices had two source and drain terminals on top, whereas a few others were made with four terminals to probe metal contact effects. Figure 3 shows microscopy images of the representative 2H-TaSe$_2$-Ti/Au devices. The majority of the devices had a channel length, $L$, in the range from 3 to 12 µm, and the channel width, $W$, in the range from 3 to 4 µm. The estimated average thickness of the devices selected for the study was $H \sim 20\text{-}35$ nm. We note that the devices with thinner channels were less robust.

[Figure 3]

### III. Electrical and Optical Characteristics of All-Metallic Devices

In Figure 4 we present source-drain I-V characteristics of 2H-TaSe$_2$ – Ti/Au all-metallic devices at RT. The curves of different color correspond to different sizes of the 2H-TaSe$_2$ channels ($L$, $W$ and $H$). The source – drain current, $I_{DS}$, was intentionally limited to a low value to avoid damaging the devices. The maximum current passed through these devices was in the µA range. One can see that all devices reproducibly reveal a well-defined threshold voltage, $V_{TH}$, for the transition from the "negligible-current" regime to the "high-current" regime. In this sense these all-metallic devices operate as switches. The value of $V_{TH}$ varied from device to device depending on the channel sizes. The high value of the source drain voltage, $V_{DS}$, which changed from -9 V to +9 V at RT, is related to the high contact resistance and imperfections of the technology used to fabricate these proof-of-concept devices from exfoliated 2H-TaSe$_2$. The insert shows I-V characteristics at T=100 K. The transition from the "Off" to "On" state is even more pronounced at low temperature.

[Figure 4]

The all-metallic 2H-TaSe$_2$ – Ti/Au devices were gated at RT. Figure 5 shows the normalized drain-source I-V characteristics for several values of the back-gate bias, V$_G$. The gate bias changes from -25 V to +25 V with the maximum V$_{DS}$=20 V. The large absolute values of the applied gate bias are explained by the fact that it was applied via thick SiO$_2$ layer (300 nm). The large V$_{DS}$ value is attributed to the voltage drop on the contacts and electrodes. The important observation for the proposed all-metallic circuits is the fact that the threshold voltage V$_{TH}$ can be shifted by the gate voltage V$_G$. The fabricated all-metallic 2H-TaSe$_2$ – Ti/Au devices can operate as gate-controlled switches. It is expected that the values of V$_G$ and V$_{DS}$ can be decreased and I$_{DS}$ increased as the device fabrication technology matures. We found that as-grown 2H-TaSe$_2$ samples had at least three orders-of-magnitude lower resistance values than the exfoliated thin film channels.

[Figure 5]

We also have studied the effect of light illumination on I-V characteristics of 2H-TaSe$_2$ – Ti/Au devices. Figure 6 shows the normalized drain-source current for the device in dark and under natural light at T=100 K. The behavior at RT was similar but the on-off transition was less pronounced. One can see that the on-set of conductive regime shifts to higher voltages in devices under illumination. This rather unusual trend is opposite from that observed in conventional photoresistors where the current increases under light illumination due to the electron – hole pair generation. However, such behavior was previously reported for CDW devices with the threshold for collective current shifted to larger V$_{TH}$ under light illumination.[26] The sharp on-off transition and strong dependence of V$_{TH}$ with light indicate some potential for optoelectronic device applications.

[Figure 6]

## IV. Discussion of Physical Properties

The physics of the process in 2H-TaSe$_2$ channel or TaSe$_2$ – Ti/Au interface, which results in the observed current characteristics, is not entirely clear. The things that can be stated now is that the effects are reproducible in many tested devices (about 20) and in the same devices over a substantial time period (about a month for some devices). The gating of the source-drain current in the metallic channel with $H\sim20$ nm is surprising due to the small carrier screening length in such materials. It is possible that the flakes have non-uniform thickness (smaller at the edges) and part of the current goes via a much thinner film allowing for gating. One should note that similar gating was observed in CDW devices with rather thick channels as well.[27] No commonly acceptable explanation exists for gating of CDW devices at RT. The measured I-V characteristics with the threshold voltage appear similar to those of CDW devices where the abrupt increase in current results from the onset of the collective current regime.[28-30] This mechanism would explain the I-Vs of 2H-TaSe$_2$ – Ti/Au devices at low temperature. However, the incommensurate CDW transition temperature, $T_P$, in 2H-TaSe$_2$ is T=122 K which is followed by a commensurate CDW transition at T=90 K.[31] Our previous work indicated that $T_P$ increases in some 2D materials with decreasing film thickness.[32] It still remains to be shown if thinning 2H-TaSe$_2$ to 20-nm thickness can result in substantial $T_P$ change.

We performed *ab initio* simulations of the band structure of bulk, bilayer and monolayer 2H-TaSe$_2$ in order to understand possible changes in electron transport. Our calculations were based on first-principles density functional theory (DFT) using the projector augmented wave method as implemented in the software package VASP.[33] The screened Heyd-Scuseria-Ernzerhof (HSE)[34] hybrid functional has been employed for this study. A Monkhorst-Pack scheme was adopted to integrate over the Brillouin zone with a k-mesh 9 x 9 x 1 (8 x 8 x 4) for the monolayer and bilayer (bulk) structure. A plane-wave basis kinetic energy cutoff of 280 eV was used. The van-der-Waal interactions in $TaSe_2$ were accounted for using a semi-empirical correction to the Kohn-Sham energies when optimizing the bulk structure.[35] The optimized lattice parameters for the bulk are a= 3.45 Å, c=13.057 Å and $z_{Se}$ = 0.131. The lattice constant for the monolayer and bilayer $TaSe_2$ structures was obtained from the optimized bulk $2H\text{-}TaSe_2$ structure. The atomic coordinates within the monolayer and the bilayer $TaSe_2$ structures were optimized by introducing a 20 Å vacuum layer between the adjacent structures. Spin-orbit coupling was included self-consistently within the band structure calculations. For the HSE calculations, 25% short-range exact Hartree-Fock exchange was used with the Perdew-Burke-Ernzerhof (PBE) correlation. The HSE screening parameter, μ, was empirically set to 0.2 (1/Angstrom). The calculated HSE electronic band structures for the normal bulk, monolayer and bilayer $TaSe_2$ structures are shown in Figure 7 (a-b). All of structures have several bands crossing the Fermi energy. Bulk $2H\text{-}TaSe_2$ is a metal. When the bulk structure is reduced to a single bilayer or monolayer, the electronic structure remains metallic. Our analysis indicates that the experimentally observed I-V characteristics are not related to metal – semiconductor phase transition.

[Figure 7]

It has been suggested that metallic $MX_2$ are more reactive and after exfoliation may undergo different chemical transformations.[36] In two of the fabricated devices we changed the metal contact technology and instead of Ti/Au deposited pure Au. The measured I-V characteristics still revealed the same non-linearity. Our experiments with bulk crystals of $2H\text{-}TaSe_2$ indicate that the resistivity of material in device channels with few-nanometer thickness is substantially higher than that in bulk form. It has also been found that I-V characteristics in bulk $2H\text{-}TaSe_2$ resistors are linear. The devices with thin exfoliated $2H\text{-}TaSe_2$ channels and different contact electrode area revealed similar characteristics for the reversed current directions, suggesting that the device channel itself is responsible for the observed phenomena.

## V. Possible Circuit Applications

The measured I-V characteristics of the all-metallic 2H-TaSe$_2$ – Ti/Au devices are completely different than those of conventional transistors. However, they can be used for information processing. The high transconductance near V$_{TH}$, strong gating effect, and low leakage current are the three unique and rather surprising characteristics of the fabricated TaSe$_2$ devices. These characteristics are unusual for the metallic-channel transistors. The measured I-V characteristics of the TaSe$_2$ devices can be fitted with the following formula

$$I_{DS} \approx I_0 \times e^{\left(\frac{V_{GS} - V_{TH}}{V_t}\right)} \times \left(1 - e^{\frac{-V_{DS}}{V_t}}\right) \times e^{\frac{\alpha V_{DS}}{V_t}}, \tag{1}$$

where $I_{DS}$ is the source-drain current, $V_{GS}$ is the gate to source voltage, $V_{DS}$ is the drain to source voltage, $V_{TH}$ is threshold voltage, $I_0$ is a constant depending on the channel geometry, $V_t$ is a parameter similar to the thermal voltage in conventional devices, $\alpha$ is a constant reflecting drain-to-channel coupling. It is interesting to note that the same equation describes the subthreshold current of an MOS device taking into account the drain-induced barrier lowering effect.[37] This effect becomes prominent in sub-micron technologies, where the source and drain depletion regions penetrate significantly into the channel and control the potential and the field inside the channel. In our case of all-metallic switches, similar I-V characteristics are observed in the long channel devices.

Another feature, which makes these devices potentially suitable for practical applications, is the experimentally observed asymmetry in the I-V characteristics for 2H-TaSe$_2$ – Ti/Au devices with different channel sizes. In Figure 8 (a) we show the experimental I-Vs for selected bias voltage that provide the functionality. The red and the blue curves correspond to the devices with two different channels: device A and device B. The threshold voltages of these experimental I-V curves are shifted by several volts. For example, the device A is in the low-conductance state in the $V_{DS}$ region from -18 V to -15 V while the device B shows exponential current increase. The situation is opposite in the voltage range from +15 V to +18 V, where device B is in the "on" state while the device A is in the "off" state. This asymmetry provides a possibility for implementing the complementary pair logic based on the TaSe$_2$ devices with different channels.

The schematics of the all-metallic inverter comprising two $TaSe_2$ devices are shown in Figure 8 (b). The devices are assumed to have a shift of the threshold voltages $\Delta V_{TH}$. The latter makes it possible to realize an all-metallic complementary pair. The same input voltage applied to the gates of these devices makes one of the transistors in the highly conducting (On state), while reducing source-to-drain current in the second transistor (Off state). As a result the whole circuit operates a NOT gate similar to the standard design with two metal-oxide-semiconductor field-effect transistors (MOSFETs). The proposed all-metallic logic gate can occupy a niche for certain applications together with conventional MOSFETs, similar to recently reported non-Boolean graphene circuits.[38]

[Figure 8]

## VI. Conclusions

We have described the fabrication and testing of all-metallic three-terminal devices with tantalum diselenide thin-film conducting channels. These devices with nanometer-scale thickness reveal strongly non-linear current-voltage characteristics, unusual optical responses and electrical gating at RT. The drain-source current in thin-film 2H-$TaSe_2$–Ti/Au devices undergoes an abrupt transition from a highly resistive to a highly conductive state, during which the threshold is tunable by the gate voltage. The combination of such current-voltage characteristics with all-metallic components may be ideal for the implementation of radiation-hard logic circuits, among other applications. These proof-of-concept results can be expanded with $TaSe_2$ thin films fabricated by chemical vapor deposition or other type of controlled film growth.

*Acknowledgements*

Work in the Balandin and Lake groups was funded by the National Science Foundation (NSF) and Semiconductor Research Corporation (SRC) Nanoelectronic Research Initiative (NRI) project "Charge-Density-Wave Computational Fabric: New State Variables and Alternative Material Implementation" NSF-1124733 as a part of the Nanoelectronics for 2020 and Beyond (NEB-2020) program. This work used the Extreme Science and Engineering Discovery

Environment (XSEDE), which is supported by the National Science Foundation grant number OCI-1053575

**FIGURE CAPTIONS**

**Figure 1:** (a) Photograph and (b) scanning electron microscopy (SEM) image of 2H-TaSe$_2$ crystals grown by chemical vapor transport. (c) Three views of the 2H-TaSe$_2$ structure (box indicates unit cell).

**Figure 2:** Raman spectra of 2H-TaSe$_2$ exfoliated films on Si/SiO$_2$ substrate. The bands at 207 – 210 cm$^{-1}$ and 233 – 235 cm$^{-1}$ correspond to E$^1_{2g}$ and A$_{1g}$, respectively. Samples were excited by the 633-nm laser in a backscattering configuration.

**Figure 3:** Optical (a) and SEM (b) images of representative all-metallic devices with 2H-TaSe$_2$ thin-film channels and Ti/Au source and drain contacts. The pseudo colors in the SEM image were used for clarity: yellow corresponds to metal contacts, green corresponds to 2H-TaSe$_2$.

**Figure 4:** Source-drain I-V characteristics of 2H-TaSe$_2$ – Ti/Au all-metallic devices at RT. The curves of different color correspond to different sizes of the 2H-TaSe$_2$ channels. Note that all devices revealed a well-defined threshold voltage, V$_{TH}$, for the transition from the "Off" to "On" state. The inset shows I-V characteristics at T=100 K.

**Figure 5:** Source-drain I-V characteristics of a representative 2H-TaSe$_2$ – Ti/Au all-metallic device for different gate biases at RT. The gate is changing from -25 V to +25 V in the direction indicated by arrows. The high absolute values of the gate bias are due to the thick SiO$_2$ layer used as gate dielectric in the back-gate design. The source-drain voltage and currents are normalized to facilitate explanation of the proposed all-metallic logic gate design.

**Figure 6:** Normalized drain-source current for the device in dark and under natural light at T=100 K. One can see that the on-set of the conductive regime shifts to higher voltages in devices under illumination, which is an opposite of what is expected for a conventional photoresistor.

**Figure 7:** *Ab-initio* band structure of 2H-TaSe2: (a) Bulk, (b) monolayer (red) and bilayer (green) thin films. The bulk band structure is plotted along the Γ-M-K-Γ-A-L-H-A symmetry points. The 1L and 2L band structures are plotted along the Γ-M-K-Γ symmetry points.

**Figure 8:** Experimental I-Vs for two TaSe2 devices A (red curve) and B (blue curve) with different channels (a). The curves show a shift of the threshold voltages $\Delta V_{TH}$ for the positive and negative bias. Schematic shows the all-metallic inverter comprising devices A and B (b). The devices are arranged in the complementary pair, where applying of the same gate voltage turns one of the switches to the conducting (on state) while reducing source-to-drain current in the second transistor to negligible (off state).

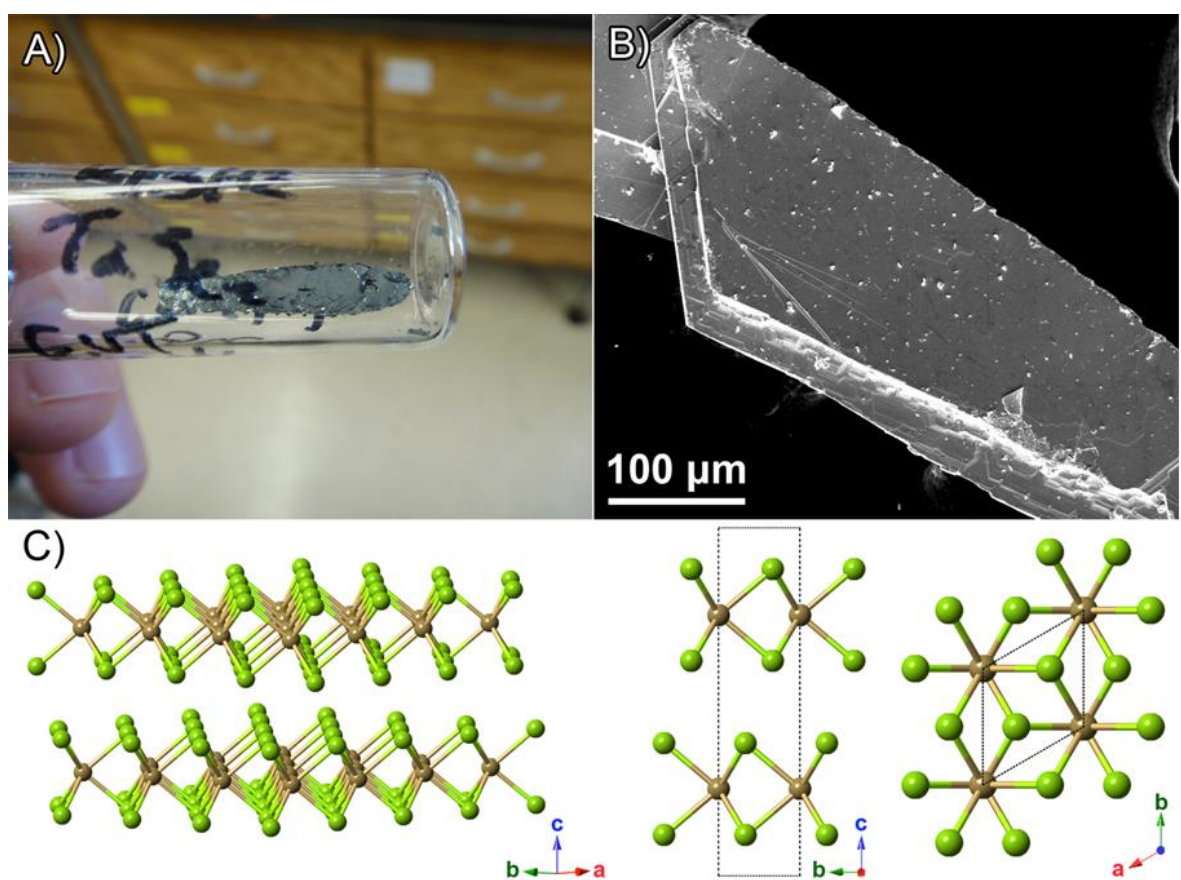

Figure 1

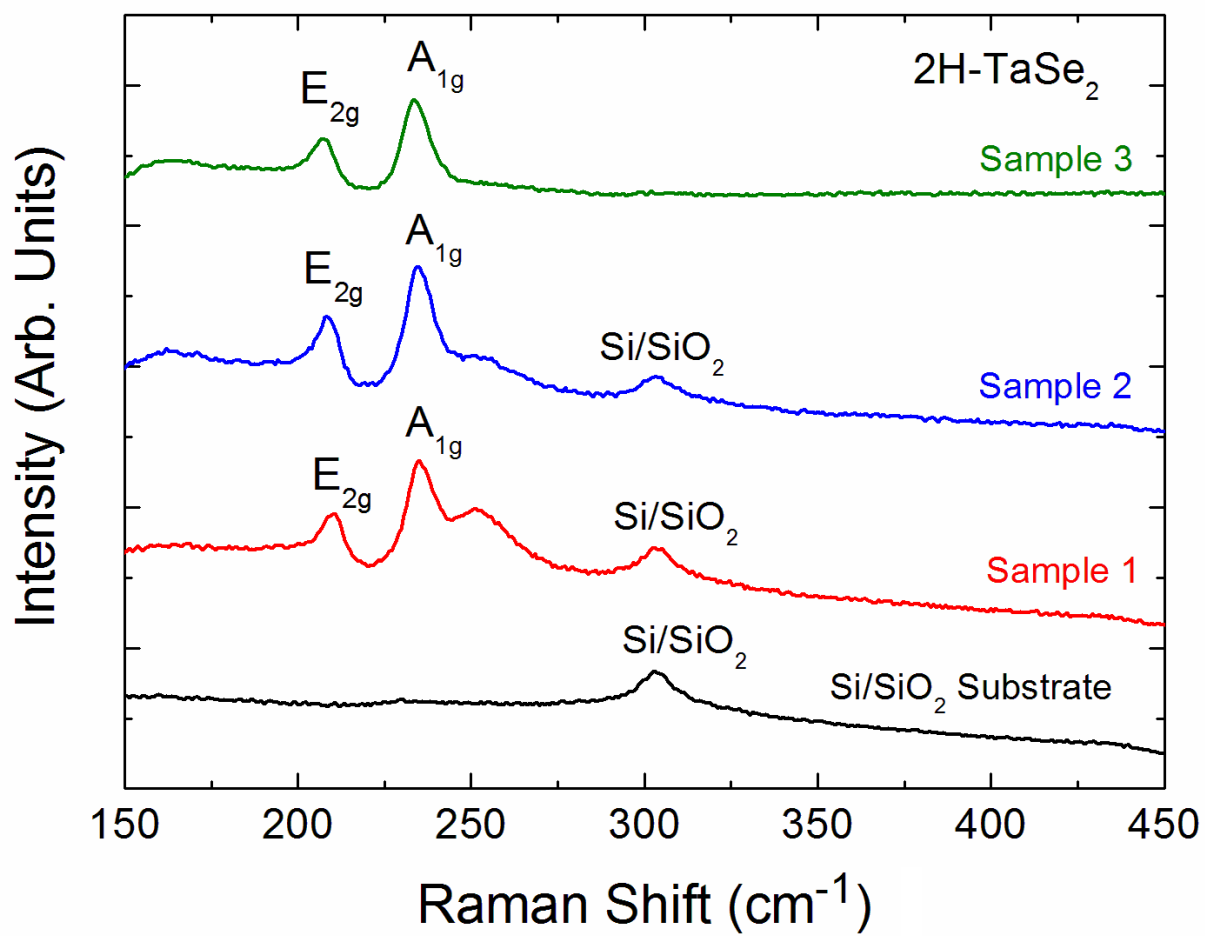

Figure 2

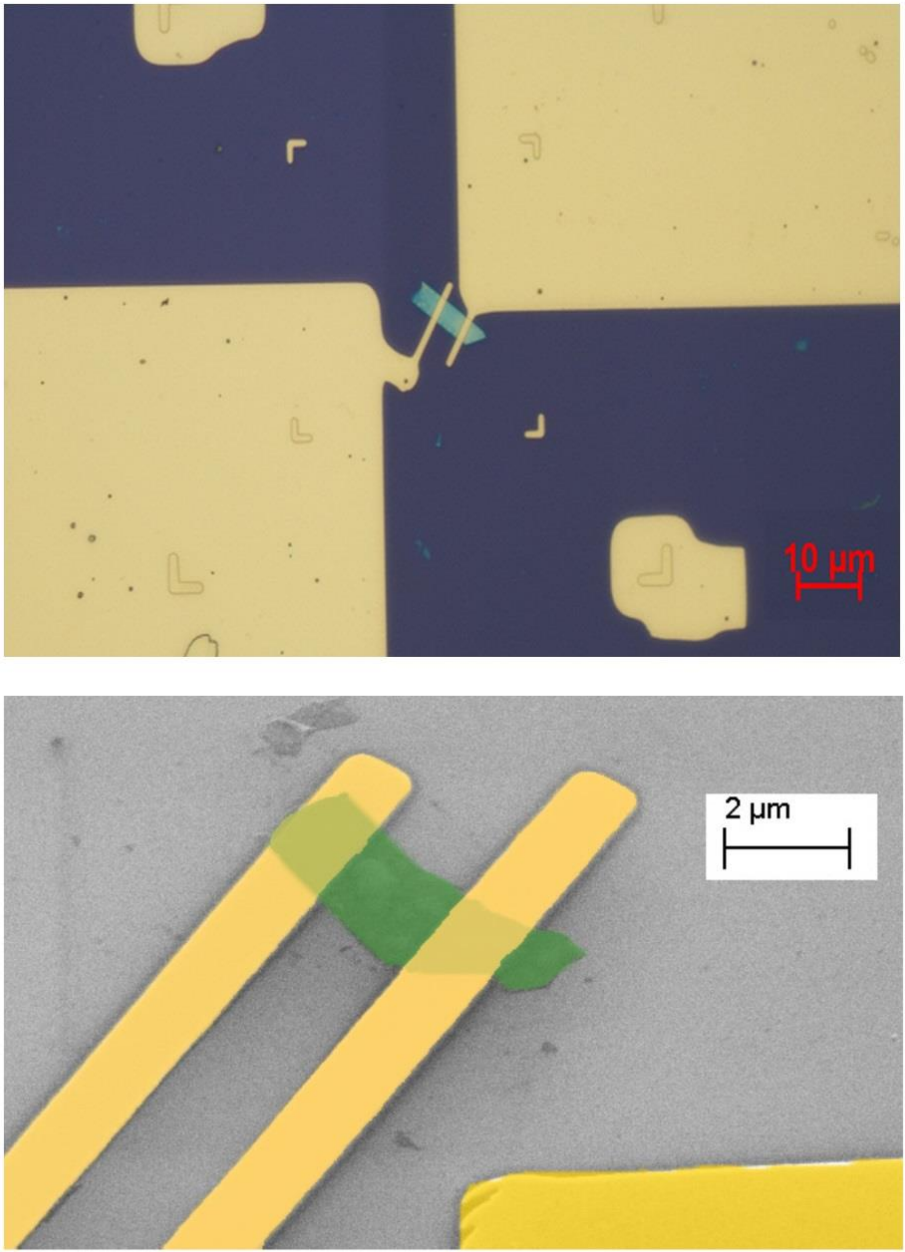

Figure 3

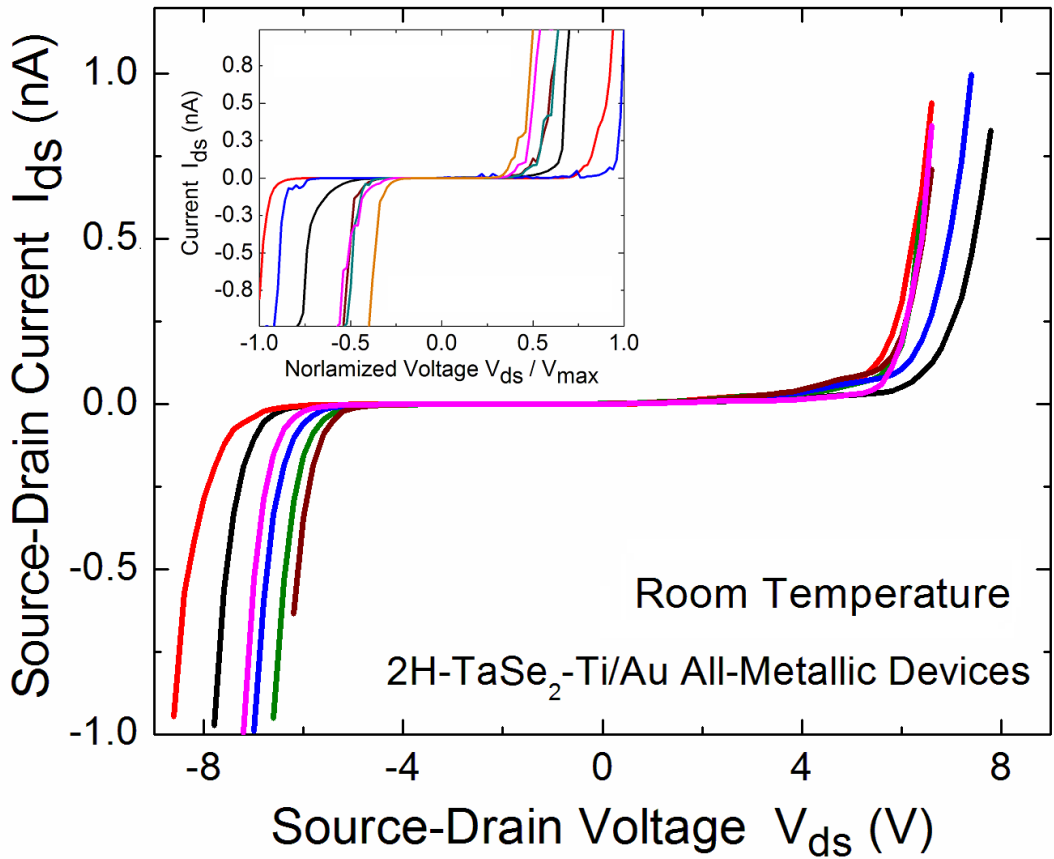

Figure 4

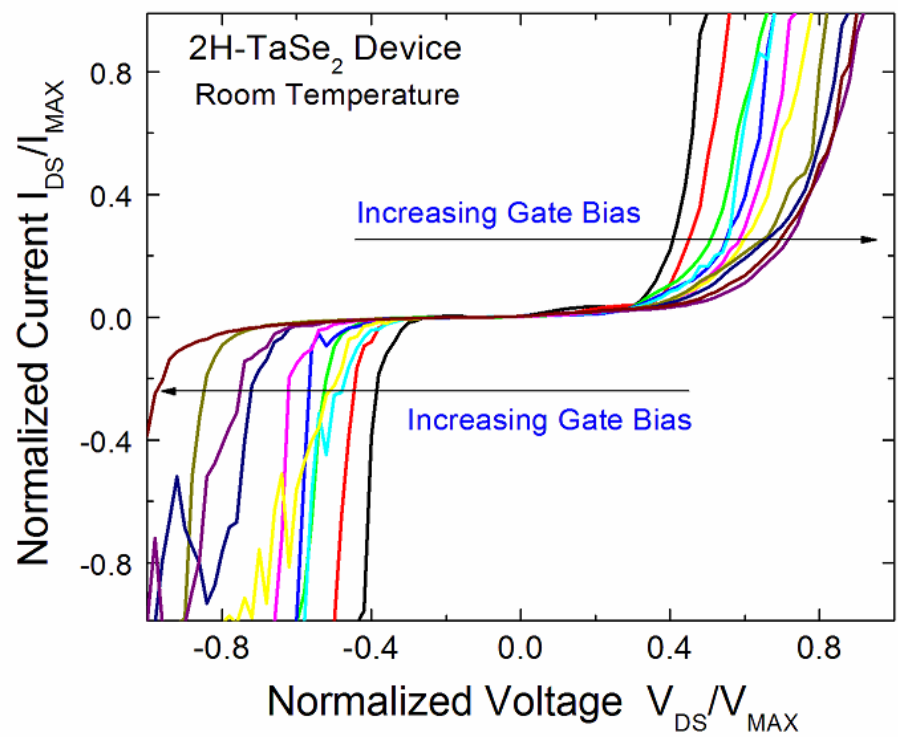

Figure 5

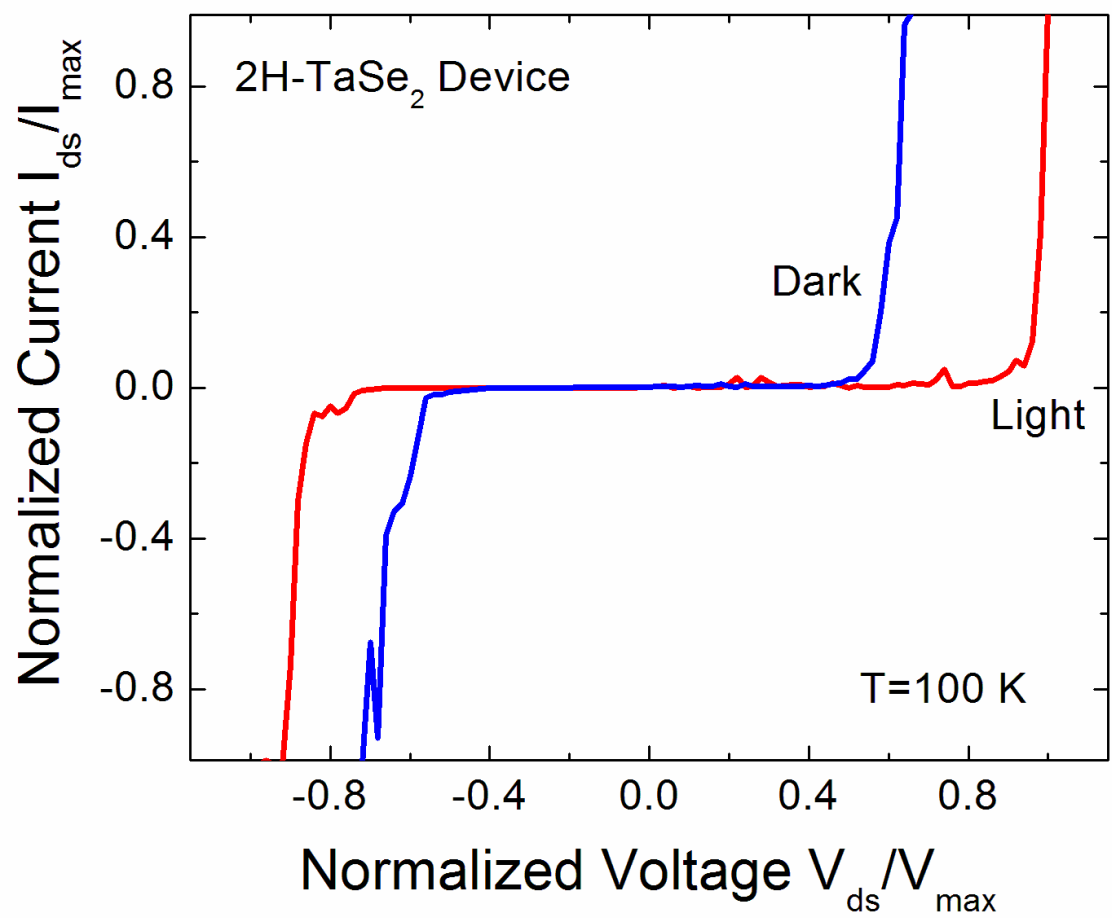

Figure 6

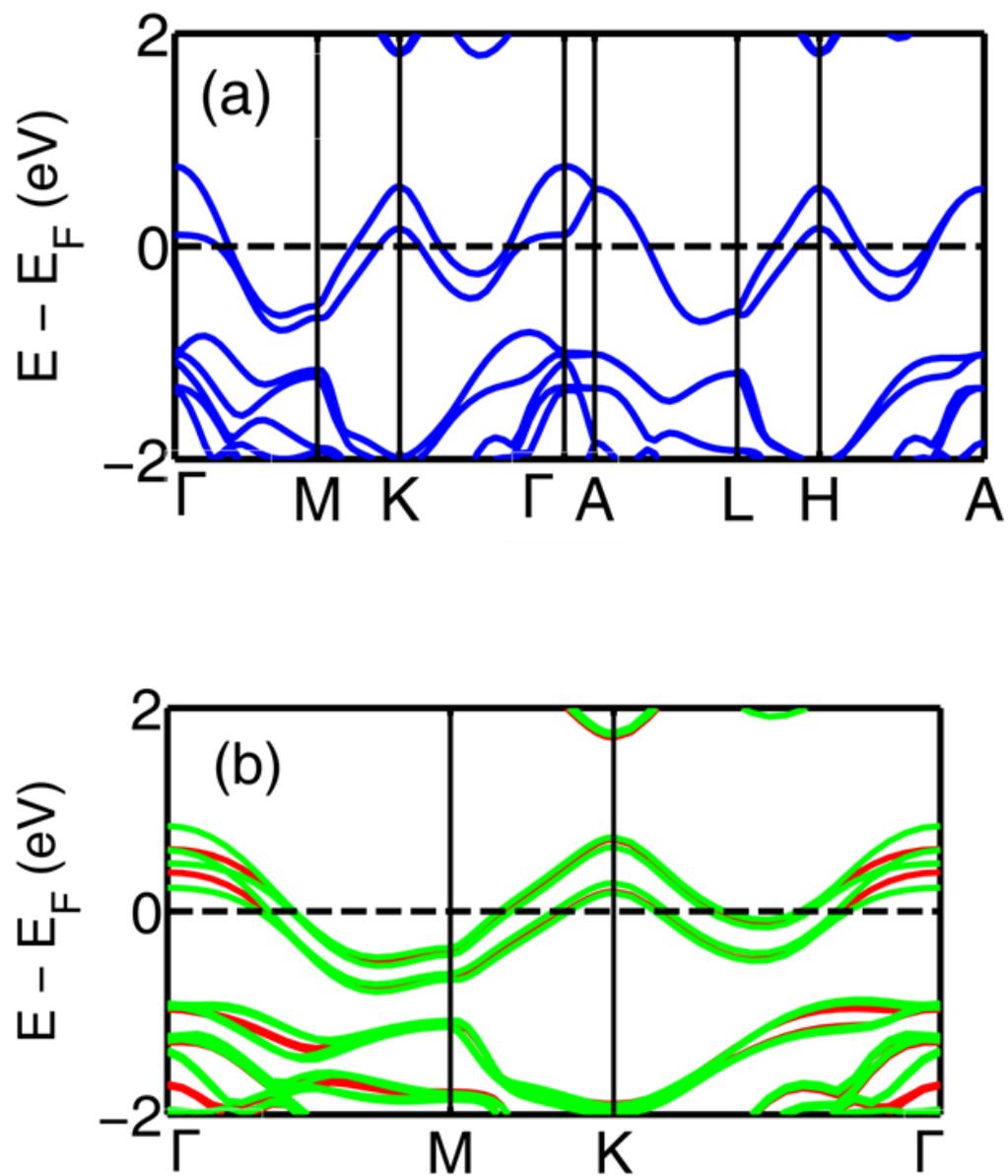

Figure 7

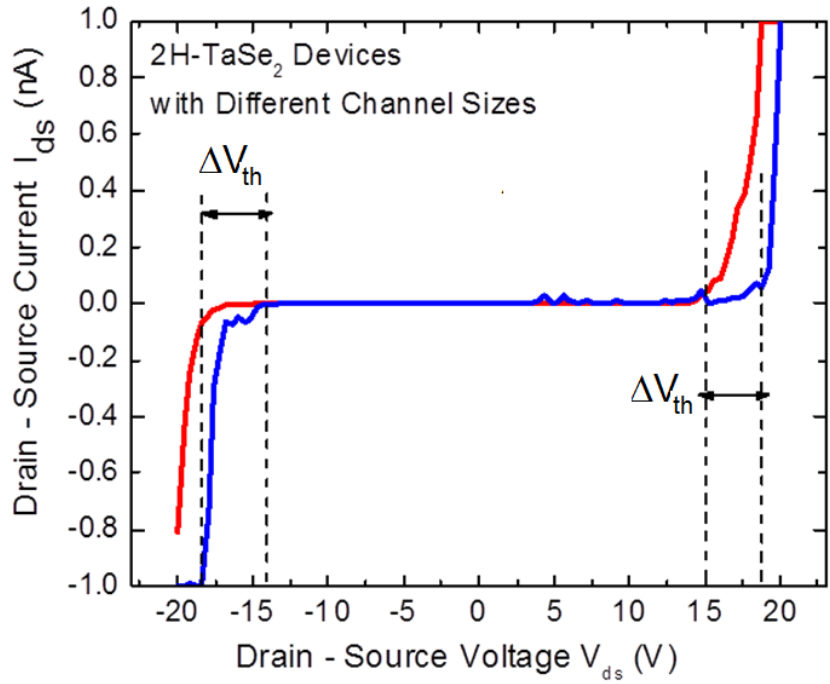

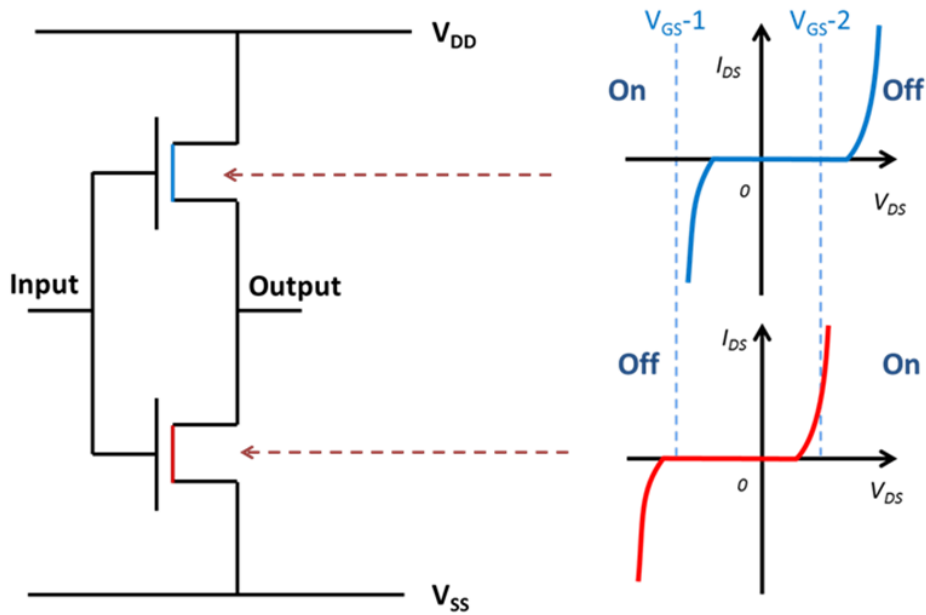

Figure 8